# Λ/Λ̄ polarization and splitting induced by rotation and magnetic field

Kun Xu,[*] Fan Lin,[†] Anping Huang,[‡] and Mei Huang[§]

*School of Nuclear Science and Technology, University of Chinese Academy of Sciences,
Beijing 100049, China*

The global polarization of $\Lambda/\bar{\Lambda}$ and the splitting between them induced by rotation and magnetic field has been investigated in a dynamical quark model by taking into account the axial-vector interaction and the anomalous magnetic moment of quarks. It is found that the rotation leads to the spin polarization of quarks and antiquarks with the same sign, while the magnetic field leads to the opposite sign, which corresponds to the $\bar{\Lambda} - \Lambda$ polarization splitting. The combination of the two effects leads to perfect agreement with experimental data. Quantitatively, the axial-vector interaction contributes 30% of the global polarization and the anomalous magnetic moment of quarks contributes 40% to the splitting of $\bar{\Lambda} - \Lambda$ polarization. However, at $\sqrt{s_{NN}} \leq 7.7$ GeV, it still remains a challenge to reach enough magnitude of the magnetic field at freeze-out.

## I. INTRODUCTION

The hot and dense matter created in noncentral heavy-ion collisions (HIC) shows the fastest rotation with angular momentum at a range of $\mathcal{O}(10^4)\hbar - \mathcal{O}(10^5)\hbar$ [1] or vorticity of order $10^{21}\text{s}^{-1}$ [2], and the strongest magnetic field $B \sim 10^{18}G$ at RHIC while $B \sim 10^{20}G$ at LHC [3]. It was firstly predicted by Liang and Wang that such fast rotation leads to global-spin polarization [4]. The polarization of $\Lambda$ and $\bar{\Lambda}$ then was observed by the STAR Collaboration in Au + Au collisions at $\sqrt{s_{NN}} = 7.7$–200 GeV [5], and the splitting of $\Lambda - \bar{\Lambda}$ polarization was also found, especially at lower energy. Recently, the polarization of $\Xi^-$ and $\Omega^-$ at 200 GeV [6] and $\Lambda$ at 3 GeV [7] were reported, which still respects the naive behavior that it increases with decreasing collision energy.

The experimental data shows two main features: (1) the global spin polarization $P_{\Lambda/\bar{\Lambda}}$ gets stronger at lower collision energy; (2) $P_{\bar{\Lambda}}$ is larger than $P_\Lambda$, and the splitting of $P_{\Lambda/\bar{\Lambda}}$ also grows with decreasing collision energy. The global polarization is well understood in the Wigner function method [8,9], the thermal model [10–14], the hydrodynamics [14–17] and the transport simulation [2,18–21], while the splitting of $P_{\Lambda/\bar{\Lambda}}$ remains as a puzzle. Possible explanations include the strong magnetic field [17,22–24], the chiral-helical vortical effect [25], and the interaction between the "magnetic" part of vector meson with hyperons [26]. However, model calculation of dynamical quarks with appropriate interactions is still missing.

In this paper, we investigate the spin polarization of dynamical quarks under rotation and magnetic field separately. We find the rotation leads to the spin polarization of quarks and antiquarks with the same sign, while the magnetic field leads to opposite sign, which corresponds to the $P_{\Lambda/\bar{\Lambda}}$ splitting. Complete results are obtained by adding them up naively. And it is found that appropriate interactions between quarks lead to a perfect quantitative agreement with experimental data.

## II. SPIN POLARIZATION INDUCED BY ROTATION AND MAGNETIC FIELD

### A. Free quark

Consider a free fermion system under slow rotation $\vec{\Omega} = \Omega\vec{z}$, and a magnetic field $\vec{B} = B\vec{z}$, the Lagrangian of which takes form of [27–33]

$$\mathcal{L}_0 = \bar\psi[\gamma^0(i\partial_t + \Omega\hat{J}_z) + i\gamma^1(\partial_x + iqBy/2) + i\gamma^2(\partial_y - iqBx/2) + i\gamma^3\partial_z - m_f + \mu\gamma^0]\psi, \quad (1)$$

where $\hat{J}_z$ is the angular momentum operator along $z$-axis $\hat{J}_z = -ix\partial_y + iy\partial_x + \frac{1}{2}\sigma^{12}$ with $\sigma^{12} = i[\gamma^1, \gamma^2]/2$. It's worthy to point out that higher-order correction from rotation, which replaces $i\partial_t$ to $i\partial_t(1 - r^2\Omega^2)$, is ignorable for the parameter set used in this paper. Similar to the quark chemical potential $\mu$, one can read that the angular velocity $\Omega$ plays the role of the spin potential.

[*]xukun21@ucas.ac.cn
[†]linfan19@mails.ucas.ac.cn
[‡]huanganping@ucas.ac.cn
[§]huangmei@ucas.ac.cn

Under external magnetic fields, the lowest Landau level (LLL) is occupied by positive(negative)-charged quark with spin (anti)parallel to magnetic field alone, while both spins take position at higher Landau levels (LLs). Then the averaged spin under magnetic field is

$$\langle s_z \rangle_B^0 = N_c \frac{qB}{2\pi} \int \frac{dp_z}{2\pi} \frac{1}{1+\exp[(E_0 \pm \mu)/T]}, \quad (2)$$

where $E_0 = \sqrt{m_f^2 + p_z^2}$ is the energy of the quark at LLL, and it's obvious that magnetic field contributes to the spin polarization of quarks and antiquarks with opposite sign.

In the slow angular-velocity case, e.g., $\Omega/T \ll 1$, which is true for RHIC, the averaged spin takes form of

$$\langle s_z \rangle_\Omega^0 = \frac{N_c}{8\pi^2} \sum_l \int dk_t^2 dk_z \frac{1}{2} [J_l^2(k_t r) - J_{l+1}^2(k_t r)]$$
$$\times \frac{1}{1+\exp[(E+\Omega(l+1/2)\pm\mu)/T]}, \quad (3)$$

where $E = \sqrt{m_f^2 + p_z^2 + p_t^2}$ and $J_l$ is the Bessel function of the first kind, and $r$ is the radius. Notice $J_0$ gives the main contribution at low momentum, then expands $\langle s_z \rangle_\Omega^0$ with respect to $\Omega/T$ and only keep $J_0$, we obtain

$$\langle s_z \rangle_\Omega^0 \approx -\frac{\Omega}{T} \int dk_t^2 dk_z f(1-f) J_0(k_t r), \quad (4)$$

where $f = 1/(1+(E\pm\mu)/T)$ is the Fermi-Dirac distribution. It indicates that the spin polarization induced by rotation is blind to the charge. This result agrees with those obtained in thermal approach [8] and Wigner function [9].

### B. Quark interaction

Until now, most methods used to study hyperon polarization were generally lack of interactions. The quarks and gluons in the fireball created in RHIC are described by strong interactions, thus it is necessary to consider the interactions between quarks to obtain realistic results.

In the general case without the magnetic field and rotation, we take four-fermion contact interaction

$$\mathcal{L}_\chi = \frac{G_S}{2} \sum_{a=0}^{8} \{(\bar\psi \lambda^a \psi)^2 + (\bar\psi i\gamma^5 \lambda^a \psi)^2\}$$
$$- G_K \{\det \bar\psi(1+\gamma^5)\psi + \det \bar\psi(1-\gamma^5)\psi\}, \quad (5)$$

the first line is (pseudo)scalar channel while the second is the 't Hooft determinant terms. $\lambda^a$ are the Gell-Mann matrices in flavor space, and $G_S$ and $G_K$ are coupling constants. This is the interaction in the three-flavor Nambu-Jona-Lasinio(NJL) model [34–36] and is widely used for the realization of spontaneous chiral symmetry breaking and restoration.

$\mathcal{L}_\chi$ preserves rotational symmetry, which is broken explicitly under the magnetic field or rotation. Thus, new interaction terms are necessary for more realistic exploration, and are supposed to be different for magnetic field and rotation considering the difference between them, which will be considered separately in the following.

#### 1. Magnetic field

In this case, $\Omega = 0$, $B \neq 0$. The tensor interaction $(\bar\psi \sigma^{\mu\nu} \psi)^2 + (\bar\psi \gamma^5 \sigma^{\mu\nu} \psi)^2$ under external magnetic fields has been investigated in [37,38], and in the mean field approximation (MFA), only the first term contributes a nonzero condensate, $\langle \bar\psi \sigma^{\mu\nu} \psi \rangle \neq 0$, which is equivalent to an anomalous magnetic moment (AMM) of quarks. Then the tensor interaction now can be presented by an AMM term coupled to the magnetic field, and the Lagrangian takes form of

$$\mathcal{L} = \bar\psi[\gamma^0 i\partial_t + i\gamma^1(\partial_x + iqBy/2) + i\gamma^2(\partial_y - iqBx/2)$$
$$+ i\gamma^3 \partial_z - m_f + \mu\gamma^0 + \kappa qB\sigma^{12}]\psi + \mathcal{L}_\chi. \quad (6)$$

The dispersion relation can be solved from corresponding Dirac equation and is $E_s^2 = p_z^2 + [\sqrt{M_f^2 + (2n+1-s\xi)|qB|} - s\kappa qB]^2$, where $\xi = \text{sgn}(qB)$ and $s = \pm 1$. $M_f$ is the effective mass; $M_f = m_f - 2G_S \sigma_f + 2G_K \sigma_{f+1} \sigma_{f+2}$, and the chiral condensates are obtained by solving the gap equations. The detailed calculation can be found in Refs. [38,39], where the induced AMM in magnetic field has been investigated carefully. Then the spin polarization induced by the magnetic field with AMM of quarks can be defined as

$$P_B = \frac{N_+ - N_-}{N_+ + N_-}, \quad N_{+/-} = N_c \frac{qB}{2\pi} \sum_n \int \frac{dp_z}{2\pi} f(E_{+/-}). \quad (7)$$

As discussed in Ref. [38,39], the AMM is supposed to be related with the chiral condensate, and we assume $\kappa = v\sigma^2$ where $v$ is treated as a free parameter to tune AMM.

#### 2. Rotation

Now $B = 0$, $\Omega \neq 0$. In general, (axial-)vector interaction should also be considered, which takes form of [40]

$$\mathcal{L}_A = -G_A\{(\bar\psi \gamma^\mu \psi)^2 + (\bar\psi \gamma^\mu \gamma^5 \psi)^2\}, \quad (8)$$

where $G_A$ is the coupling constant. Under mean field approximation, the first term contributes to nothing but the chemical potential and is ignored in this work. And it is not hard to find that the operator $\bar\psi i\gamma^0 \gamma^1 \gamma^2 \psi$ corresponds to the number(density) difference of spin-up and spin-down, and also notice $\gamma^3\gamma^5 = i\gamma^0\gamma^1\gamma^2$, thus it is reasonable to take $\mathcal{L}_A = -G_A(\bar{\psi}\gamma^3\gamma^5\psi)^2$. It is instructive to notice that in Wigner function method, the mean spin vector is proportional to the integral of the axial-vector component of the covariant Wigner function over some arbitrary 3D spacelike hypersurface [41].

Similar to the chiral condensate $\sigma_f = \langle\bar{\psi}_f\psi_f\rangle$, we define the axial-vector condensate $a_f = \langle\bar{\psi}_f\gamma^3\gamma^5\psi_f\rangle$, then the Lagrangian under MFA takes the form of

$$\mathcal{L}_{\text{MFA}} = -G_S\sum_f \sigma_f^2 + 4G_K\sigma_u\sigma_d\sigma_s + G_A\sum_f a_f^2$$
$$+ \bar{\psi}[\gamma^0(i\partial_t + \Omega\hat{J}_z) + i\vec{\gamma}\cdot\vec{\partial} - M_f + \mu\gamma^0 + \mu_S\gamma^3\gamma^5]\psi,$$
(9)

where $\mu_S = -2G_A a_f$ is an effective spin potential. Then the quark field can be solved from corresponding Dirac equation and the general solution of positive energy reads

$$\psi(\theta, r) = e^{-Et+ik_z z}\begin{pmatrix} ce^{il\theta}J_l(k_t r) \\ ide^{i(l+1)\theta}J_{l+1}(k_t r) \\ c'e^{il\theta}J_l(k_t r) \\ id'e^{i(l+1)\theta}J_{l+1}(k_t r) \end{pmatrix}, \quad (10)$$

where $J_l(k_t r)$ is the Bessel function of the first kind. $c$, $d$, $c'$, and $d'$, are coefficients, and satisfies $c^2 + d^2 + c'^2 + d'^2 = 1$ required by normalization, and $d/c = -d'/c' = \lambda$, where we define

$$\lambda = 2\mu_S k_t /\left\{\left[E_{l,s} + \left(l + \frac{1}{2}\right)\Omega - \mu_S\right]^2 - M_f^2 - k_t^2 - k_z^2\right\},$$
(11)

and the dispersion relation now is

$$E_{l,s}(k_t, k_z) = \sqrt{(\sqrt{M_f^2 + k_z^2} - s\mu_S)^2 + k_t^2} - \left(l + \frac{1}{2}\right)\Omega,$$
(12)

with $l$ the quantum number of angular momentum. $s = \pm 1$, but does not correspond to spin(or helicity) as in the case without spin potential [27]. The negative-energy solution can be obtained in similar way.

Now the grand potential under MFA is

$$V = \int rv(r)dr, \quad (13)$$

where $v(r)$ is the "grand potential density" at radius $r$. Due to the rotation, all the condensates are supposed to be functions of radius $r$, however, to simplify the calculation, we made an assumption that all the condensates, thus effective mass and spin potential, change slowly with radius, and such assumption also can be found in [19,28]. The grand potential density now has form of

$$v(r) = G_S\sum_f \sigma_f^2 - 4G_K\sigma_u\sigma_d\sigma_s - G_A\sum_f a_f^2 - \sumint_R E_{l,s}$$
$$- T\sumint_R\left\{\ln\left(1 + e^{-\frac{E_{l,s}+\mu}{T}}\right) + \ln\left(1 + e^{-\frac{E_{l,s}-\mu}{T}}\right)\right\},$$
(14)

where we have made the abbreviation

$$\sumint_R = \frac{N_c}{8\pi^2}\sum_f\sum_{l=-\infty}^{+\infty}\sum_{s=\pm}\int dk_t^2 dk_z W_{l,s}(k_t, k_z), \quad (15)$$

with $W_{l,s}(k_t, k_z) = [J_l^2(k_t r) + \lambda^2 J_{l+1}^2(k_t r)]/(1 + \lambda^2)$. It is clear from $W_{l,s}(k_t, k_z)$ that the energy eigenstate has a definite total angular momentum (TAM) $J = l + 1/2$, with two combinations of orbital angular momentum (OAM) and spin angular momentum (SAM); $(l, 1/2)$ and $(l+1, -1/2)$. The spin-orbit coupling is realized through the function $W_{l,s}(k_t, k_z)$.

Assume $m_u = m_d = m_l$, $\sigma_u = \sigma_d = \sigma_l$, $a_u = a_d = a_l$, then the equilibrium state can be obtained by solving the functional gap equations

$$\frac{\delta V}{\delta\sigma_{l/s}(r)} = 0, \qquad \frac{\delta V}{\delta a_{l/s}(r)} = 0. \quad (16)$$

With the solution of all condensates (quark mass and spin potential) we then can calculate the quark-spin polarization. The two eigenstates of spin degenerates with nonzero spin potential, and the averaged spin of a given energy eigenstate is

$$\langle s_z\rangle_\Omega = \frac{1}{2}\left[\frac{1}{1+\lambda^2}J_l^2(k_t r) - \frac{\lambda^2}{1+\lambda^2}J_{l+1}^2(k_t r)\right]. \quad (17)$$

It is worthy of pointing out that this formula does not depend on charge, e.g., spin polarization for fermion and antifermion has same sign. Now the spin polarization induced by rotation including the axial-vector interaction can be defined as

$$P_\Omega = \frac{1}{Z}\frac{N_c}{8\pi^2}\sum_{l,s}\int dk_t^2 dk_z\langle s_z\rangle_\Omega f(E_{l,s}), \quad (18)$$

where $Z$ is a local normalization constant,

$$Z = \frac{N_c}{8\pi^2}\sum_{l,s}\int dk_t^2 dk_z f(E_{l,s}). \quad (19)$$

## III. NUMERICAL CALCULATION

Several parameters need to be fixed in the numerical calculations, and by fitting to the pion mass, kaon mass, and pion decay constant in vacuum, we take $m_l = 0.0055$ GeV, $m_s = 0.136$ GeV, $\Lambda_p = 0.627$ GeV, $G_S = 3.282/\Lambda_p^2$, and $G_K = 14.244/\Lambda_p^5$. Besides, a regularization scheme is necessary since the NJL model is non-normalizable, and we take

$$\int d^3\mathbf{p} E(\mathbf{p}) \to \int d^3\mathbf{p} E(\mathbf{p}) \frac{\Lambda_p^{2N}}{\Lambda_p^{2N} + (\mathbf{p}^2)^N}, \quad N = 15. \quad (20)$$

Notice that if a Fierz transformation is performed to scalar interaction, the coupling constants of the scalar and axial-vector channels are equal (with a minus sign) [34,37]; thus, it is reasonable to set $G_A = -G_S$.

By solving gap equations [Eq. (16)] we obtain the chiral and axial-vector condensates and thus the quark effective mass $M$ and spin potential $\mu_S$. As discussed above, the condensates (as well as thermal quantities) are supposed to change with radius; however, it is found numerically that at the region of $r < 1$ GeV$^{-1}$, the quantities, for example, quark mass and spin polarization, do not change significantly with the radius; Thus we take $r = 0.1$ GeV$^{-1}$ in the following, which is about 0.02 fm.

The effective mass $M$ and spin potential $\mu_S$ of $u/d$ quarks as functions of temperature are shown in Fig. 1. Now the chiral symmetry is still restored at high temperatures; however, the quark effective mass as well as the critical temperature are suppressed by rotation. For large angular velocity, for example, $\Omega = 0.15$ GeV, the phase transition changes from crossover to first order, and similar behavior was also observed in Ref. [27]. Similar to that, the magnetic field induces AMM and angular velocity induces spin potential. At vanishing angular velocity $\Omega = 0$, zero $\mu_S$ is obtained, while nonzero angular velocity leads to nonzero $\mu_S$, and with increasing angular velocity, $\mu_S$ increases consequently. Besides, $\mu_S$ also increases with temperature, and saturates above phase transition temperature. Spin potential will lead to larger spin polarization compared to the case with zero $\mu_S$.

Next, to make a quantitative comparison with the experimental data, we consider the spin polarization along the freeze-out line in the HIC. Now the baryon chemical potential, temperature, angular velocity, and magnetic field at freeze-out are treated as input and are supposed to be extracted from experimental data. For baryon chemical potential and temperature, there are already fitted functions [36,42]; $\mu_B = 1.477/(1 + 0.343\sqrt{s_{NN}})$ and $T = 0.158 - 0.14\mu_B^2 - 0.04\mu_B^4$, where $\sqrt{s_{NN}}$ is the collision energy. The angular velocity can also be estimated from the hyperon polarization in the thermal model of free fermions [5]. However, there are about 80% hyperons emitted directly from the fireball, which contributes to the final hyperon polarization; thus, we take $\Omega_{\exp} \approx 0.8T(P_\Lambda + P_{\bar\Lambda})$. For the magnetic field at freeze-out, we extract $eB \approx \frac{M_N T}{0.613}(P_{\bar\Lambda} - P_\Lambda)$ [43], where $M_N = 0.938$ GeV.

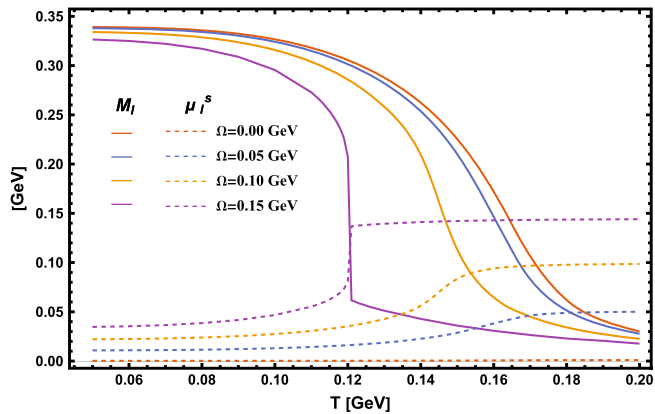

FIG. 1. The effective mass $M$ and spin chemical potential $\mu_S$ of $u/d$ quark as functions of the temperature and angular velocity under vanishing baryon chemical potential.

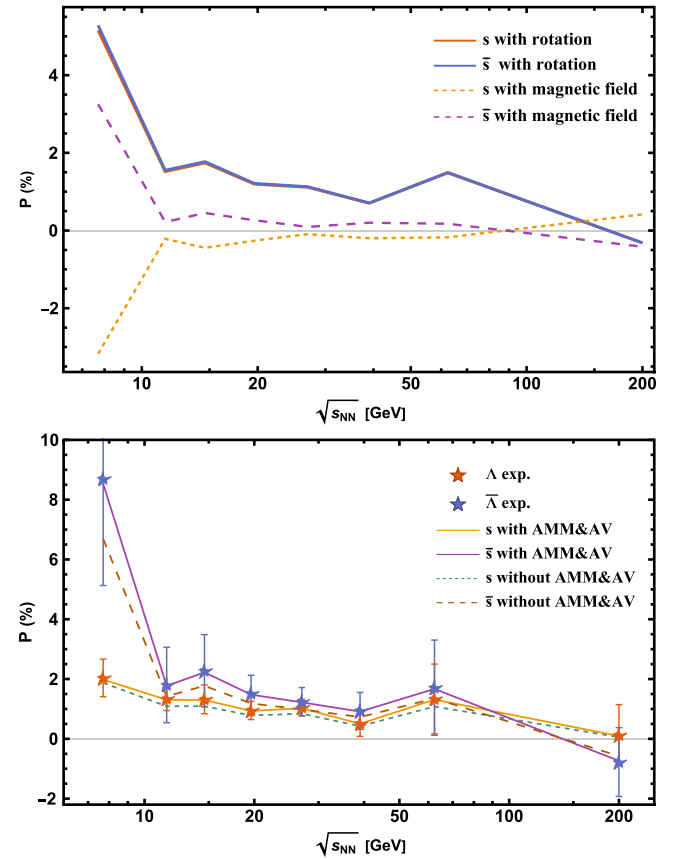

FIG. 2. (Top panel) The spin polarization of $s/\bar s$ induced from rotation and magnetic field separately. (Bottom panel) The total spin polarization of $s/\bar s$ with and without the contribution from axial vector interaction and AMM, respectively. Experimental data are from [5,6] for $\Lambda/\bar\Lambda$ polarization.

With $\Omega$ and $eB$ extracted from global polarization and the splitting of polarization based on the thermal model of free fermion system, we then use the dynamical quark model (including axial-vector interaction and AMM to rederive) the polarization of $s/\bar{s}$, and the results are shown in Fig. 2. Here the hyperon-spin polarization is calculated in a simple quark-coalescence model, where the spin polarization of $\Lambda/\bar{\Lambda}$ is from $s/\bar{s}$ quark [4,44]. It is clearly seen that rotation leads to the same polarization for $s$ and $\bar{s}$, while the magnetic field gives positive polarization for $\bar{s}$ and negative for $s$. This explains why the polarization for $\bar{\Lambda}$ is larger than that of $\Lambda$. By adding them naively as shown in the bottom panel of Fig. 2, we found that the rotation leads to the global polarization of $\Lambda/\bar{\Lambda}$, while the magnetic field leads to the splitting. It is also found numerically that the axial-vector interaction contributes an extra 30% of the global polarization.

For the magnitude of the magnetic field to be reached at freeze-out for noncentral HIC, we consider two sources: (1) The remaining magnetic field from the evolution of the system after collision, which includes the contributions of spectators and the response of the QGP, and we fit the results of [45] from the space-average magnetic field at time $t = 5$ fm, and (2) The magnetic field induced by vorticity [46], in which a strong magnetic field along the fluid vorticity is naturally generated by the charged subatomic swirl. We use the formula $eB = \alpha_s nA\Omega$, herein $\alpha_s = e^2/4\pi$, $A = \pi R^2$, and $n = 0.30 - 0.087 \ln\sqrt{S_{NN}} + 0.0067(\ln\sqrt{S_{NN}})^2$. In our calculations, we take the radius $R = 4$ fm, and the vorticity $\Omega$ used in Fig. 2. The polarization difference $P_{\bar{\Lambda}} - P_{\Lambda}$ induced by these two sources of magnetic field are shown in Fig. 3. It shows that the splitting $P_{\bar{\Lambda}} - P_{\Lambda}$ agrees well with experimental data except at $\sqrt{s_{NN}} = 7.7$ GeV, where the magnetic field obtained with these two sources is too small and the induced polarization splitting is around 1.5%, which is much smaller than the experimental data. This indicates that we need to reanalyze the evolution of the magnetic field at low-collision energies where the vorticity is large, thus the vorticity-induced magnetic field should be taken into account simultaneously. Furthermore, we want to emphasize that the AMM of quarks contributes 40% to the splitting of $\bar{\Lambda} - \Lambda$ polarization.

## IV. SUMMARY AND OUTLOOK

In the present paper we investigate the quark spin polarization in a magnetic field and a rotating frame separately. Qualitative analysis from the free case shows that the rotation leads to spin polarization to quarks and antiquarks with the same sign, while the magnetic field gives opposite sign. To make realistic results we consider the interactions in the three-flavor NJL model and take into account the axial-vector interaction and the AMM of quarks, which is generally lacking in widely used methods. Besides, the conditions of freeze-out, temperature, baryon chemical potential, angular velocity, and magnetic field, are extracted from experiments.

Assuming that $P_{\Lambda/\bar{\Lambda}}$ is equal to $P_{s/\bar{s}}$, we find our model results are in perfect agreement with experimental data; the rotation leads to the increase of $P_{\Lambda/\bar{\Lambda}}$ with collision energy decreases, while the magnetic field induces $\bar{\Lambda} - \Lambda$ splitting. The axial-vector interaction and the AMM of quarks are of importance for the quantitative agreement; the axial-vector interaction contributes an extra 30% of the global polarization while the AMM of quarks contributes 40% to the $\bar{\Lambda} - \Lambda$ splitting.

Furthermore, we consider $\bar{\Lambda} - \Lambda$ splitting induced by the magnetic field from two sources; the remaining magnitude at freeze-out [45] and the magnetic field induced from vorticity [46]. It is found that the splitting $P_{\bar{\Lambda}} - P_{\Lambda}$ agrees well with experimental data except at the collision energy 7.7 GeV, where the magnetic field obtained with these two sources is at least 20 times smaller. This indicates that we need to reanalyze the evolution of the magnetic field at low-collision energies where the vorticity is large thus the decay of the initial magnetic field and the vorticity-induced magnetic field should be taken into account simultaneously, which may induce a larger magnitude of magnetic field at freeze-out for low-energy collisions. Besides, we assumed that quantities change slowly with radius to obtain analytical results. A full result of spin polarization can be obtained by integration over all radii, which could be more realistic compared to experiments.

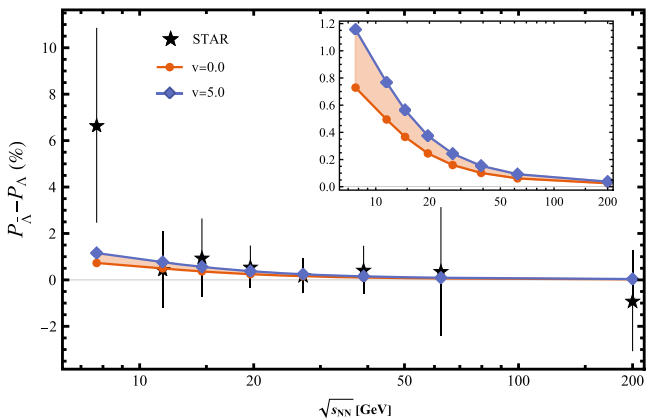

FIG. 3. The polarization difference $P_{\bar{\Lambda}} - P_{\Lambda}$ induced by the magnetic field with two sources: the remaining magnitude at freeze-out [45] and the magnetic field induced from vorticity [46], the contribution of AMM with $\kappa = v\sigma^2$ is considered. The results are compared with STAR measurement in [5,6].

## ACKNOWLEDGMENTS

This work is supported in part by the Strategic Priority Research Program of Chinese Academy of Sciences Grants No. XDB34030000 and No. XDPB15, and supported by the National Natural Science Foundation of China (NSFC) Grants No. 12147150, No. 11725523, No. 11735007, No. 12235016, No. 12221005, and No. 12205309, and the start-up funding from University of Chinese Academy of Sciences (UCAS), and the Fundamental Research Funds for the Central Universities.